\documentclass[aps,prl,epsfig]{revtex4}
\usepackage{graphicx}
\usepackage{amsfonts}
\usepackage{amssymb}
\usepackage{amsmath}
\usepackage{color}

\begin{document}

\title{Vortex Lines Distribution in Inhomogeneous Lattices}

\author{Mauro Iazzi}
\affiliation{SISSA, Via Bonomea 265, I-34136 Trieste, Italy}
\author{Nicola Bassan}
\affiliation{SISSA, Via Bonomea 265, I-34136 Trieste, Italy}
\affiliation{INFN, Sezione di Trieste, I-34127 Trieste, Italy}
\author{Andrea Trombettoni}
\affiliation{SISSA, Via Bonomea 265, I-34136 Trieste, Italy}
\affiliation{INFN, Sezione di Trieste, I-34127 Trieste, Italy}
\author{Kevin E. Schmidt}
\affiliation{Department of Physics, Arizona State University, Tempe, AZ 85287, USA}
\author{Stefano Fantoni}
\affiliation{On leave from SISSA, Via Bonomea 265, I-34136 Trieste, Italy}
\affiliation{ANVUR, National Agency for the Evaluation of Universities and Research Institutes, P.za Kennedy 20, I-00144 Roma, Italy}

\begin{abstract}
We study the distribution of vortex lines in a three-dimensional lattice 
with inhomogeneous couplings. We investigate  
the spatial distribution of the number of vortex lines, showing how 
the vortex lines are expelled from the region with higher couplings. 
Results for the correlation functions are presented, 
together with a discussion of the 
implications of the obtained results for 
the vortex distribution in ultracold trapped gases and in neutron stars.
\end{abstract}

\maketitle

\section{Introduction}
\label{intro}

The study of vortex and vortex lines in superfluids and superconductors 
is an evergreen field of research \cite{tilley90,annett04,leggett06}. 
In type-I superconductors, the rigidity of the wavefunction of superconducting electrons 
determines the expulsion of the magnetic field (Meissner effect): the superconductivity 
is destroyed at a critical magnetic field. However, when the value 
of the Ginzburg-Landau parameter exceeds a critical value, then the magnetic field can 
enter the superconductor: the flux penetrates in a regular array 
of flux tubes, each carrying a flux quantum \cite{degennes89,tinkham96}. The corresponding properties 
of vortices in type-II superconductors have been then intensively studied in BCS and high-temperature 
superconductors \cite{tinkham96,blatter94,brandt95,rosenstein10}

Vortices have been extensively studied also in neutral superfluids: as it is well known, 
the effect of a rotation on neutral particles is equivalent to that of a magnetic field on a system of charged 
particles \cite{leggett06}. As a consequence, one has in neutral superfluids the counterpart of the effects regarding 
the penetration of a magnetic field in a superconductor: the expulsion of the magnetic field has its equivalent in that a neutral superfluid, up to a certain value of the rotation, does not rotate together with a rotating 
trap or vessel. Thus the use of Ginzburg-Landau theory to describe vortex states is not restricted to superconductivity and 
it can be applied, with certain adjustments, to neutral superfluids.

The study of vortex states in neutral superfluids witnessed in the last two decades a substantial additional interest 
due to progress in the realization, control and manipulation of ultracold bosonic and fermionic gases 
\cite{pitaevskii03,pethick08}. Correspondingly, the study of properties of vortices in ultracold gases 
subjected to rotating external potentials attracted a significant attention \cite{pitaevskii03,pethick08,fetter09}. 
For trapped Bose-Einstein condensates, when the condensate rotates with angular velocity $\Omega$
higher than a critical value, one or several vortices nucleate: 
for more rapid rotations, 
the vortices form a dense triangular array \cite{fetter09}. At even higher rotating frequencies the superfluid 
properties are lost and when $\Omega$ is close to the radial trap frequency the lowest-Landau-level approximation 
becomes applicable \cite{fetter09}: the rotating dilute gas is expected to exhibit a transition 
from a superfluid to various strongly correlated, non-superfluid states analogous to those 
typical in the fractional quantum Hall effect for electrons in a strong perpendicular magnetic field 
\cite{fetter09,cooper08,viefers08}. 

We also mention that the recent experimental realization of 
synthetic magnetic and electric fields acting on neutral atoms 
using a suitable spatial variation and time dependence of the effective
vector potential \cite{lin09,lin11} opens the way to simulate 
many-body systems in controllable (static) electro-magnetic fields 
\cite{bloch08}: such a possibility, 
eventually together with the use of rotating traps 
\cite{fetter09,cooper08}, opens the way to the study of bosonic and/or 
fermionic systems in a vast class of simulated magnetic fields 
\cite{dalibard10}, including spin-orbit coupling.

The study of vortices has been carried out both in two- and in three- 
dimensional systems. In two dimensions, the thermally generated 
vortices form, at low temperature, pairs of opposite vorticity, giving 
rise to power-law correlation functions. At the critical 
point of the corresponding Berezinskii-Kosterlitz-Thouless (BKT) 
transition the vortices start to unbind and high-temperature 
exponential decaying correlations arise \cite{berezinskii71,kosterlitz73}. 
A paradigmatic model in which the 
BKT transition has been deeply investigated is the $XY$ model on a lattice, 
which is also 
commonly used to describe defect-mediated phase transitions and superfluidity 
in two dimensions \cite{nelson02}. The unbinding of the vortices around the $XY$ transition 
point was studied and visualized using Monte Carlo simulations \cite{chester79}. 
The $XY$ model description has the 
advantage that it is possible to extract the topological degrees of freedom 
(i.e., the vortices) by constructing the dual theory \cite{savit80}, 
showing that the vortices interact via a logarithmic potential. 
A physical realization of the lattice $XY$ model is provided by 
Josephson junction arrays, i.e. superconducting grains located at the vertices 
of a network \cite{martinoli00,fazio01}. Furthermore, for large 
number of particles per site and negligible fluctuations of the number, 
the effective Hamiltonian of ultracold bosons in optical lattices can be 
mapped to that of an (eventually quantum) $XY$ model 
\cite{polini05,trombettoni05,kasamatsu09}: the vortex number and 
the BKT transition were experimentally detected 
in a two-dimensional Bose gas in an optical lattice at finite 
temperature \cite{schweikhard07}.

Passing from two to three dimensions, e.g. by coupling two-dimensional systems, 
then vortex lines connect the core of the two-dimensional vortices: 
the vortex lines must either close on themselves (vortex rings) or terminate 
on boundaries \cite{leggett06}. Following the original ideas put forward 
by Onsager \cite{onsager49} and Feynman \cite{feynman55} that 
vortex excitations are responsible for the superfluid transition in $^4He$, 
the properties of vortices in three-dimensional superfluid systems were 
subsequently investigated: a real-space renormalization technique was used 
in \cite{williams87} to generate screened vortex energies and core sizes. 
The $XY$ model is useful also to study vortex excitations in three dimensions: 
using the dual description, it is possible to write the Hamiltonian 
in terms of vortex loop degrees of freedom \cite{savit80}. Monte Carlo results 
highlighted the role of vortex loops, showing 
indeed that they are responsible for the phase transition in the three-dimensional $XY$ model 
\cite{kohring86}. Using a 
three-dimensional generalization of the two-dimensional 
Kosterlitz real-space scaling procedure, Shenoy wrote down 
scaling equations for the vortex-loop fugacity and the loop coupling 
\cite{shenoy89}. This computation was then generalized to the anisotropic 
three-dimensional $XY$ model \cite{biplab94} in order to study 
the $3D$--$2D$ crossover.

Important contributions to theory of vortices and vortex lines 
were made by Luciano Reatto and his collaborators 
\cite{chester68,reatto93,sadd96,sadd97}: in \cite{chester68} 
the energy of a $^4He$ vortex line and the density distribution in the 
core region were computed using model many-body wavefunctions. 
The velocity of translation of the vortex ring was also computed and 
shown to be in agreement 
with experimental results. The variational energy was calculated using the 
Percus-Yevick and the convoluted-hypernetted-chain integral equations: the 
structure factor entering these equations was taken by measured values and 
assumed linear, with slope $\hbar k/2 mc$, at small wavevectors 
\cite{chester68} --- as it is well known, a model wavefunction leading 
to this result was constructed in the seminal paper \cite{chester67}. 
The properties of vortex lines in $^4He$ were reconsidered in 
\cite{sadd96,sadd97} where a shadow wavefunction was used 
(optimized shadow wavefunctions for the ground state of liquid and solid 
$^4He$ and at the liquid-solid $^4He$ interface 
were studied in \cite{pederiva95,moroni96,moroni98}). 
In \cite{sadd97} the excitation energy of the vortex 
was then computed showing a very weak dependence 
on the density when the latter is increased from equilibrium to near 
freezing, giving insights on the properties of the classical core parameter 
\cite{sadd97}.

The problem we address in the present paper is the study of the effects 
of inhomogeneities on the vortex lines 
distribution in three-dimensional lattices. As motivation of our work, 
we have in mind mostly two systems: 
trapped ultracold atoms in optical lattices and neutron stars. Regarding the 
former, the low-energy properties of ultracold atoms in deep optical lattices 
are in general described by Hubbard-like models both for bosons 
\cite{jaksch98} and fermions \cite{hofstetter02}. When the system is 
superfluid, in each well $i$ of the periodic potential one can define 
a phase $\phi_i$: either due to temperature and/or to the effect of a rotation, 
vortex lines will develop in the system. Since an external potential is 
usually present in such trapped systems \cite{pitaevskii03,pethick08}, 
then an arising issue is the determination of how the trapping potential 
affects the distribution vortex lines.

The study of the effect of inhomogeneities on the vortex lines in superfluids in a 
lattice is also relevant in the context of neutron stars (NS). NS are compact objects, 
with typical masses of order $1.5$ solar masses and  typical radii of order $10\,km$,  
formed from the collapsed core of fairly high mass stars when they explode as supernovae. 
Apart from their extreme compactness, observed NS are usually characterized by very high rotation frequencies --- up to a $kHz$ --- 
and very high magnetic fields --- up to $10^{14}$--$10^{15}\,G$.
A NS can be roughly subdivided into three zones having different composition and properties: 
the first is a shell called the outer crust composed of a lattice of atomic nuclei permeated by an electron gas. 
The outer crust ends at about $100\,m$ from the surface, where the density reaches $10^{11}\,g\,cm^{-3}$. 
The next layer in is called the inner crust and consists of a lattice of very neutron rich nuclei permeated by free neutrons and electrons. 
Theoretical calculations \cite{fabrocini08,gandolfi09} and indirect observational evidence \cite{shternin10,page10} support the idea that 
crustal free neutrons form a superfluid. The space dependence of the superfluid gap has been also studied \cite{fabrocini08,gandolfi09}. 
The crust ends at about $1\,km$ 
from the surface where the density reaches $2.7 \times 10^{14}\,g\,cm^{-3}= 0.16\,fm^{-3}$. 
The inner region, called the core, is formed by free neutrons, protons and leptons (neglecting any more exotic composition). 
Again observations seem to suggest that neutrons are superfluid while protons are superconducting \cite{shternin10,page10}.

Vortices are expected to appear in both the inner crust and the core of a NS. Moreover, they are expected to 
interact in the crust with normal phase lattice nuclei \cite{avogadro08}. This microscopical interaction is widely believed \cite{chamel08} 
to be the mechanism that gives rise to glitches. A glitch is a sudden increase of the otherwise slowly and steady declining neutron star spin period 
(for a recent statistical study of glitches see \cite{espinoza10}).
%\footnote{A NS is expected to irradiate energy while spinning due to its magnetic axis being displaced with respect to the rotation one (Pacini mechanism). For further details see \cite{Shapiro,chamel08}}. 
The inhomogeneous neutron superfluid fills a layer spanning three 
order of magnitude in density 
and it feels an effective periodic potential created by the crustal lattice: 
free neutrons interact with the lattice nuclei feeling a one-body  
potential that can be thought as having a Thomas-Fermi like shape 
\cite{wong04}. At  
different densities both the lattice nuclei and the lattice length are  
different \cite{negele73,Pastore2011}:  
in any case the potential  
becomes negligible (because of the fine range of the nuclear force)  
without superimposing with those of the first neighbours, creating in this way a (deep) lattice potential for the superfluid neutrons.
An interesting issue is then the study of the effect of the inhomogeneity on the vortex lines in the inner crust of a rotating NS and possibly the individuation 
of a simplified model capturing the main properties of vortex lines in NS. As a first step towards 
the individuation of such a simplified model (having parameters derived from the microscopic Hamiltonian), we study in the following  
the distribution of thermally excited vortex lines for a superfluid in a lattice with inhomogeneous couplings proportional to the superfluid gap 
in the NS inner crust: for this purpose 
we use the data for the gap presented in Ref. \cite{fabrocini08}.

To explore the main properties of vortex lines in inhomogeneous lattices 
and have results independent of
the microscopic details, we decided to investigate the inhomogeneous 
version of the $XY$ model described by the Hamiltonian
\begin{equation}
H=-\sum_{\langle ij \rangle} J_{ij} \cos{\left( \phi_i-\phi_j \right)}.
\label{inh_ham}
\end{equation} 
In Eq.(\ref{inh_ham}), the sum is on the nearest-neighbours pairs of 
sites $i$, $j$ of a three-dimensional lattice (supposed cubic for simplicity) 
and $\phi_{i}$ is a phase variables, having values between $0$ and 
$2\pi$. Furthermore, $J_{ij}$ is the coupling between the sites 
$i$ and $j$, that we will assume to be inhomogeneous in the space (typically 
gaussian in one or all directions). We will mostly consider in the following 
open boundary conditions.

Several reasons lead us to consider Hamiltonian (\ref{inh_ham}) to model 
the effect of inhomogeneities on vortex lines: first, in the homogeneous limit 
($J_{ij}=J$ across the lattice), the $XY$ model can be expressed in terms 
of the topological degrees of freedom in two and three dimensions, giving 
important insight on vortex-vortex interactions and on the behaviour 
of correlation functions. Moreover, 
the Hamiltonian (\ref{inh_ham}) 
could be realized on its own by using Josephson junction arrays with 
Josephson energies $J_{ij}$ varying across the lattice or by considering 
ultracold atoms in an optical lattice, whose Hamiltonian can be mapped 
in a suitable range of parameters to (\ref{inh_ham}). 
As a last reason, more important 
for our purposes, we observe that in the presence of a superfluid in a lattice, 
the macroscopic wavefunction $\psi_j$ can be written as $\psi_j=\sqrt{n_j} 
e^{i \phi_j}$: then the Hamiltonian will have a term 
$\propto -2t\sum_{\langle ij \rangle} 
\sqrt{n_i n_j} \cos{\left(\phi_i-\phi_j\right)}$, 
where $t$ is the hopping parameter. 
This means that we can proceed with the identification 
\begin{equation}
J_{ij} \equiv 2 t \sqrt{n_i n_j}.
\label{J_ij}
\end{equation} 
Then, if we have a superfluid in which the number $n_i$ 
is fixed, e.g. by an external potential, and we want to focus on the 
equilibrium and dynamical properties due to the behaviour the phases, 
we can stick to the Hamiltonian (\ref{inh_ham}) with the identification 
(\ref{J_ij}). 

The plan of the paper is then the following: in Section \ref{homog} we remind 
the main properties of the Hamiltonian (\ref{inh_ham}) in the homogeneous limit 
($J_{ij}=J$) and we recall how the vortex degrees of freedom are introduced. 
The effect of inhomogeneities is addressed in Section \ref{inhomog}, where we 
present numerical results for the spatial correlations and 
the number of vortex lines, pointing out that  
the vortex lines are expelled from the region with large couplings. 
Results with an external magnetic field are also presented. 
Our conclusions, 
together with a discussion of perspectives and future work, are presented 
in Section \ref{concl}.

\section{The homogeneous limit}
\label{homog}

In this Section we recall the main properties of the 
Hamiltonian (\ref{inh_ham}) in the homogeneous limit, i.e. when 
the couplings $J_{ij}$ are taken equal to some value $J$, 
so that (\ref{inh_ham}) reads
\begin{equation}
H=-J \sum_{\langle ij \rangle} \cos{\left( \phi_i-\phi_j \right)}.
\label{hom_ham}
\end{equation}
We observe (\ref{hom_ham}) can be written as 
$H=-J\sum_{\langle ij \rangle} \vec{S}_i \cdot \vec{S}_j$, where 
$\vec{S}_i=\left( \cos{\phi_i}, \sin{\phi_i}\right)$. A 
detailed discussion of several properties of the three-dimensional 
$XY$ model is in \cite{kleinert89}.

The $XY$ Hamiltonian (\ref{hom_ham}) displays a phase transitions, belonging 
to the $3D-XY$ universality class at a critical temperature $T_c$ in which 
the order parameter $\langle \cos{\phi_i} \rangle$ becomes non-vanishing: 
accurate Monte Carlo estimates give \cite{janke90,gottlob93}
\begin{equation}
k_B T_c \simeq 2.202J,
\label{crit_homog}
\end{equation}
i.e. $\beta_C J \simeq 0.454$ (where $\beta_C=1/k_B T_c$).

A simple estimate of the $T_c$ can be obtained by 
mean-field \cite{schneider00}: the mean-field Hamiltonian reads 
\begin{equation}
H_{MF} = - z m J \sum_i \cos{\phi_i},
\label{H_MF}
\end{equation}
where $z=6$ is the number of nearest-neighbours and 
$m=\langle \cos{\phi_i} \rangle$ is the parameter to be 
self-consistently determined. One has
\begin{equation}
m=\frac{\int_{-\pi}^{\pi} d\phi \, \cos{\phi} \, e^{\beta z J m \cos{\phi}}}
{\int_{-\pi}^{\pi} d\phi \, e^{\beta z J m \cos{\phi}}}:
\label{m_MF}
\end{equation}
it follows
\begin{equation}
m=\frac{I_1\left( \beta z J m\right)}{I_0\left( \beta z J m\right)},
\label{m_MF_2}
\end{equation}
where $I_n\left( x \right)$ is the $n$-th modified Bessel function 
\cite{abramowitz64}. By performing the limit $m \to 0$, one finds 
the mean-field critical temperature
\begin{equation}
k_B T_c^{(M.F.)} = 3J,
\label{crit_homog_MF}
\end{equation}
which should be compared with (\ref{crit_homog}).

A useful way to go beyond mean-field results, both for the critical 
temperatures and the critical exponents, is provided by rewriting 
Hamiltonian (\ref{hom_ham}) in terms of the correct topological excitations, 
the vortex loops. To this aim, one has to resort to dual transformations 
\cite{savit80}: details are found in 
\cite{shenoy89,biplab94,biplab93,biplab_PHD}, here we sketch 
the main steps in order to show how vortex loops emerge 
in the effective description of the system.

The dual transform procedure is the same as in the two-dimensional 
$XY$ model: the main point consists in expanding $e^{-\beta H}$ in terms 
of the basis eigenfunctions invariant with respect to the symmetry 
of the Hamiltonian \cite{savit80}. It is convenient to write 
the Hamiltonian (\ref{hom_ham}) as 
\begin{equation}
H=-J \sum_{j,\mu} \cos{\Delta_\mu\phi_j},
\label{re-wr}
\end{equation}
where $\Delta_\mu$ is the discrete derivative in the direction 
$\mu$, with $\mu=\hat{x}, 
\hat{y}, \hat{z}$ denoting the versors in the three directions 
(e.g., $\Delta_{\hat{x}}\phi_j=\cos{\phi_{j+\hat{x}}}-\cos{\phi_j}$, 
and so on). To write the Hamiltonian (\ref{re-wr}) in term of the dual variables, which turn out to be 
the vortex lines themselves, one has to introduce the lattice bond 
variables $n_{j,\mu}$ (which are {\em integer} variables) and then integrate over the initial 
variables $\phi_i$ \cite{savit80}. Expanding the Boltzmann factor $e^{-\beta H}$ in terms of  
$e^{i \sum_j n_{j,\mu}\Delta_\mu \phi_j}$ and integrating over the $\phi_i$ one gets a 
constraint, as in any dual transformation \cite{savit80}: 
the constraint to be satisfied is $\sum_n \Delta_\mu n_{j,\mu} \equiv 
\vec{\Delta} \cdot \vec{n}=0$. One has then to write $n_{j,\mu}$ as the curl 
of a dual lattice bond variable in order to satisfy the 
constraint: using the Poisson summation formula \cite{savit80} amounts then to introduce the vortex loop variables 
$\vec{J}\left(\vec{r}\right)$. This procedure is the equivalent of the duality transformation that 
in $2D$ defines the Coulomb gas Hamiltonian, where the charges of the Coulomb gas are the ($2D$) vortices \cite{savit80}. 
One finally gets for the partition function
\begin{equation}
Z = \sum_{\{\vec{J}\}} e^{-\beta {\cal H}}:
\label{zeta_J}
\end{equation}
the effective 
Hamiltonian is written in terms of the vortex loop variables $\vec{J}\left(\vec{r}\right)$ and it reads  
\begin{equation}
{\cal H} = \frac{\pi K_0}{2} \sum_{\vec{r} \neq \vec{r}'} 
\vec{J}\left(\vec{r}\right) \cdot \vec{J}\left(\vec{r}'\right) 
\tilde{U}\left(\vec{r}-\vec{r}'\right),
\label{HAM_J}
\end{equation}
where $\tilde{U}\left(\vec{r}\right)$ 
is the three-dimensional lattice Green's function $U\left(\vec{r}\right)$ 
(apart from a constant): 
$\tilde{U}\left(\vec{r}\right) \equiv U\left(\vec{r}\right) - 
U \left( 0 \right)$, where 
for large distances it is $U\left( \vec{r} \right) 
\sim 1/\mid \vec{r} \mid$ \cite{kleinert89}. In Eq.(\ref{HAM_J}) $K_0 \simeq 
J$ at low temperatures (the full dependence 
upon temperature is discussed in \cite{kleinert89}).

The vortex loop variables $\vec{J}\left(\vec{r}\right)$ are vectors, having the three components
$J_\mu\left(\vec{r}\right)$ ($\mu=\hat{x}, 
\hat{y}, \hat{z}$). The components $J_\mu\left(\vec{r}\right)$ in each site $\vec{r}$ of the dual lattice are integer variables
$$J_\mu\left(\vec{r}\right)=0, \pm 1, \pm 2, \cdots:$$
at low temperatures 
only the values $J_\mu\left(\vec{r}\right)=0,\pm 1$ are relevant. One has the following 
local conservation law:
\begin{equation}
\vec{\Delta} \cdot \vec{J} \left( \vec{r} \right)=0,
\label{COND_J}
\end{equation}
meaning nothing but that the loops are closed. The physical meaning of the variables $J_\mu\left(\vec{r}\right)$ is that 
a vortex passes from one elementary cell cube of center $\vec{r}$ to a neighbour cube of center $\vec{r}+\hat{\mu}$ 
if the sum of the phase differences ($mod \, 2\pi$) across the square face separating the two cells has modulus $2\pi$, i.e. if a vortex 
line passes through the two centers (from $\vec{r}$ to $\vec{r}+\hat{\mu}$ if $J_\mu\left(\vec{r}\right)=1$, and  
from $\vec{r}+\hat{\mu}$ to $\vec{r}$ if $J_\mu\left(\vec{r}\right)=-1$). In the $2D$ case the vector $\vec{J}\left(\vec{r}\right)$ is a scalar, 
meaning that one has a counter-clockwise (clockwise) vortex around the plaquette center $\vec{r}$ if the value is $1$ ($-1$), i.e. one finds 
back the standard definition of $2D$ vortices. 
 
An important consequence of rewriting the Hamiltonian in terms of the correct degrees of freedom $J_\mu\left(\vec{r}\right)$ is that 
it is possible, starting from (\ref{HAM_J}), to perform a real-space renormalization study: scaling equations 
are obtained for the vortex-loop fugacity $y(\ell)$ and the coupling ${\cal K}(\ell)$ as a function of the scale $\ell$ 
\cite{shenoy89}. The final result is
\begin{equation}
\frac{d{\cal K}}{d\ell}={\cal K}-Ay{\cal K}^2; \, \, \, \frac{dy}{d\ell}=\left(6-\pi {\cal K} {\cal L}\right) y
\label{scaling}
\end{equation}
where $A=4 \pi^2/3$ and ${\cal L} \approx  1 + \ln{\left[a/a_c \left( \ell \right)\right]}$, with $a=e^\ell$ and $a_c\left( \ell \right)$ the finite 
core size at scale $\ell$. Analysing Eqs.(\ref{scaling}), one can have a good estimate of the critical temperature and of the critical 
exponents \cite{shenoy89}. 

The emerging physical picture is the following: the three-dimensional $XY$ model has vortex loops as topological excitations, with loop segments 
interacting via a $1/r$ potential. Thermally created loops are few and small at low temperatures: when the temperature 
is increased the loops expand and smaller loops wedge between larger loops. This screening weakens the binding of larger loops, until 
the dominating largest loops are expelled at the critical temperature, with a proliferation of vortex loops.

\section{Effect of the inhomogeneity \& numerical results}
\label{inhomog}

In this Section we study the Hamiltonian (\ref{inh_ham}) with inhomogeneous couplings $J_{ij}$: to fix the notation, we consider 
in the following three kinds of inhomogeneity. First, we consider a gaussian radially symmetric coupling given by 
\begin{equation}
J_{ij}=J_0 e^{-r^2/2R^2},
\label{J_1}
\end{equation}
where $r$ is the distance (hereafter measured in units of the lattice length) 
from a chosen origin of the midpoint of the sites $i$, $j$; $J_0$ is an overall coupling and $R$ is a length quantifying  
the spatial extent of the couplings $J_{ij}$. The choice (\ref{J_1}) refers to the physical situations in which an external 
potential forces the superfluid to stay close to the centre, as it is the case, e.g., for trapped superfluids in parabolic potentials.

Another possible choice amounts to considering the inhomogeneity only in one direction: we set 
\begin{equation}
J_{ij}=J_0 e^{-z^2/2 R^2},
\label{J_2}
\end{equation}
where $z$ is the distance from a chosen origin along the $\hat{z}$ direction of the midpoint of $i$ and $j$. The couplings 
(\ref{J_2}) correspond, in an ultracold atom setup, to a trap squeezed in the $\hat{z}$ direction and refer to a superfluid 
occupying a certain ``slice'' of a three-dimensional system: this is the case also for the neutron superfluid in the crust of a NS. 
For this reason, we also consider as a third choice couplings $J_{ij}$ proportional to the superfluid gap in the NS inner crust: we use 
the data for the gap in Ref. \cite{fabrocini08} obtained using the $f_6$ correlated model (see table $1$ in \cite{fabrocini08}). 
Since in a NS the critical temperature is order of $MeV$ and the temperature $T$ is much smaller than $T_c$, the relevant 
regime of temperatures to model NS through the Hamiltonian (\ref{inh_ham}) is the low-temperature one.   
Denoting with $\hat{z}$ the radial direction of the NS, the data from \cite{fabrocini08} 
as a function of the distance $z$ from the centre of the inner crust are well 
fitted with a function $\left(A+Bz\right)\,e^{-Cz^2}$. We will use then (with $J_0=RB$ and $z_0=RA/J_0$)
\begin{equation}
J_{ij} =  J_0 \, \frac{z}{R} \, e^{-(z-z_0)^2/2R^2}.
\label{J_3}
\end{equation}
It should be noted that even if the XY model is usually formulated with two energy scales $J$ and $T$, only their ratio has a physical meaning. Therefore all of the results only depend on the adimensional parameter $\beta J_0$ (or $\beta J$ in the homogeneous case).

As a starting point of the following discussion, 
we observe that in the homogeneous limit 
[i.e., $J_{ij}=J$ in Eq.(\ref{inh_ham})] the critical temperature $T_c$ is proportional to $J$: for temperatures smaller than $T_c$ a few small vortex loops are thermally generated. If there is a shallow dependence of the couplings, one can consider that there is a space dependent value of the critical temperature for the vortex loop blow out. Then, with the couplings varying across the lattice, as a consequence of the inhomogeneity, in the regions with small (large) coupling constants there are many (few) large (small) vortex loops. This can be rephrased in terms of a local density approximation (LDA), where the effective local temperature is $T_{\mbox{\small eff}}(\vec{r})=J_0T/J(\vec{r})$. According to this picture, there should be many vortex loops in the regions in which the coupling is very small, e.g., for large $r$ with the couplings (\ref{J_1}): however, there are no superfluid particles there, as shown by Eq.(\ref{J_ij}), and an effective number of vortices will reach a maximum value at intermediate values of the couplings. One can think to the presence of regions with a maximum effective number of vortices as related to the occurrence of vortex-vortex interactions that are no longer translationally invariant and effectively attractive in the regions with intermediate values of the couplings. One can quantitatively test the validity of the LDA by computing the local observables of interest in the homogeneous case for various temperatures and comparing the results with the local values of the same observables in the inhomogeneous system, with coupling $\beta J(\vec{r})$ as we show later in Fig. \ref{fig9}.

To quantitatively test and visualize the described scenario involving the effective exclusion of vortex loops 
from the regions with higher couplings, we numerically studied the Hamiltonian (\ref{inh_ham}) with 
the couplings (\ref{J_1})-(\ref{J_3}) by means of a Monte Carlo simulation with open
boundary conditions. Vortex lines were identified using the standard definition \cite{kleinert89}: a vortex passes from one elementary cell cube to a neighbour cell if the sum of the phase differences ($mod \, 2\pi$) across the square face separating the two cells
has modulus $2\pi$.  In other words, if the face between two cubes has the four sites $1, \dots, 4$ and phases 
$\phi_1, \dots, \phi_4$, then one computes the quantity $(1/2\pi) \sum_\alpha \tilde{\Delta}_\alpha\phi$ where 
$\alpha=1,\dots,4$ and $\tilde{\Delta}_\alpha\phi=\phi_{\alpha+1}-\phi_\alpha (mod \, 2\pi)$ (with $\phi_5 \equiv \phi_1$). 
The winding direction of the vortex line depends upon the sign of this sum (not shown in the figures).

\begin{figure}
\begin{center}
\includegraphics[angle=0,width=0.7\textwidth]{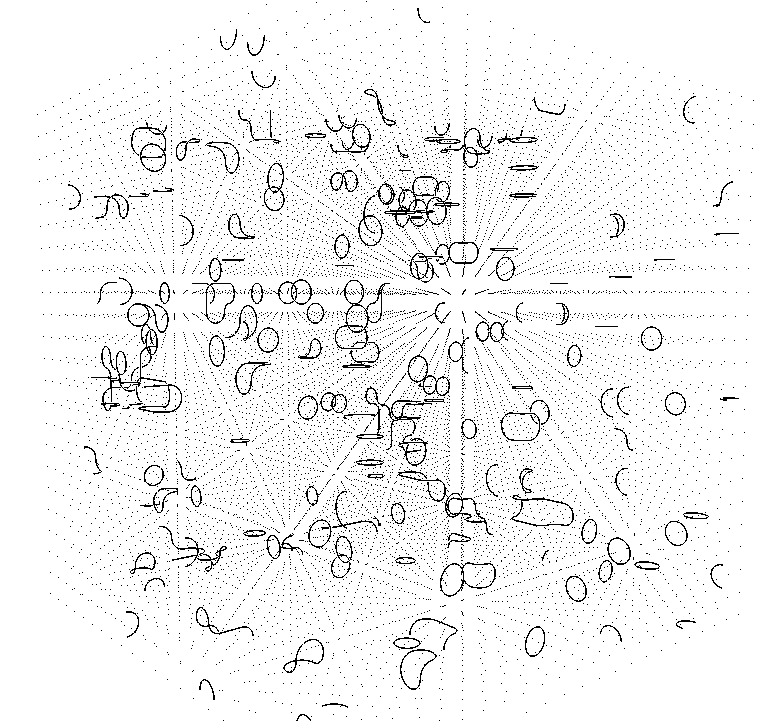}
\caption
{A Monte Carlo snapshot in the homogeneous limit for a temperature below the critical temperature ($\beta J=0.6$, 
while $\beta_C J \simeq 0.454$). The vortices are small in radius and clearly separated. In this figure and in the followings 
the sites of the cubic lattice are depicted as points in a three-dimensional image: the vortex lines, 
connecting the centres of the elementary cell faces, 
are drawn as solid lines \cite{iazzi}. %and the different colours of the vortices 
%are a visual guide for the eye to better distinguish the lines. 
}
\label{fig1}
\end{center}
\end{figure}

\begin{figure}
\begin{center}
\includegraphics[angle=0,width=0.7\textwidth]{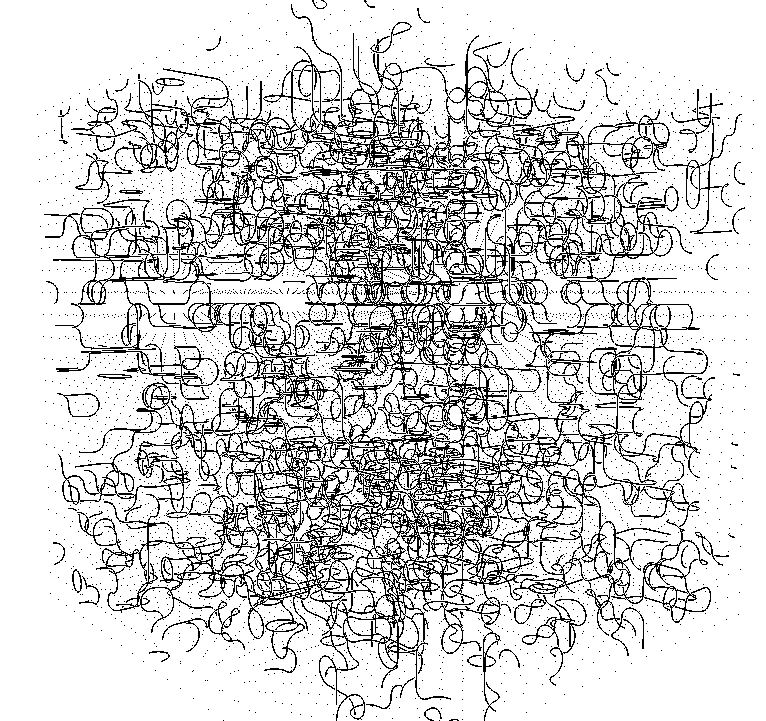}
\caption{A typical configuration above the critical temperature 
in the homogeneous limit ($\beta J=0.4$). 
Here the vortices are not separated and the inter-vortex distance is 
smaller than their radius.}
\label{fig2}
\end{center}
\end{figure}

\begin{figure}
\begin{center}
\includegraphics[angle=0,width=0.7\textwidth]{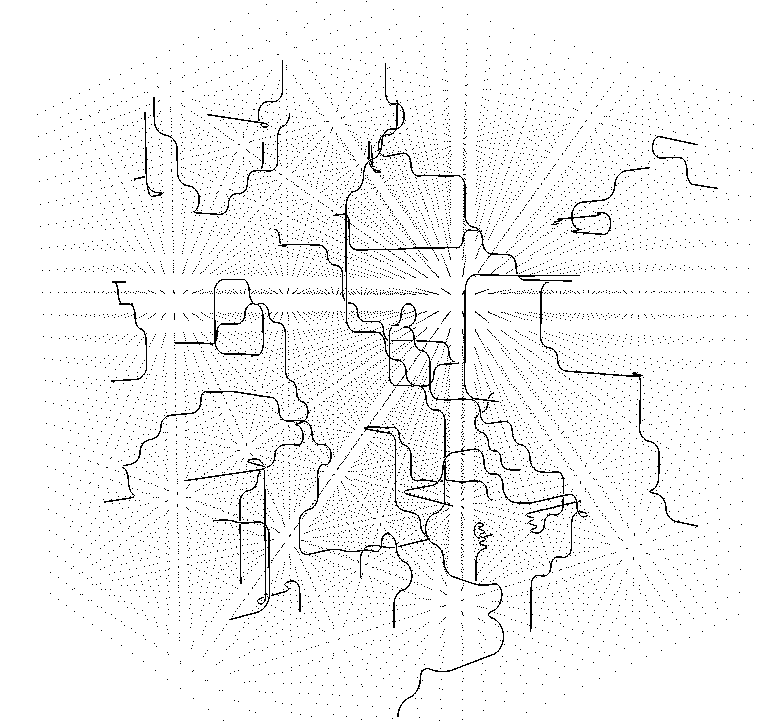}
\caption{A configuration with open vortex lines in the homogeneous limit, obtained after a quench from $T=\infty$ to $T=0.1$. 
This configuration has a high energy and is not stable (below $T_c$).}
\label{fig3}
\end{center}
\end{figure}

\begin{figure}
\begin{center}
\includegraphics[angle=0,width=0.7\textwidth]{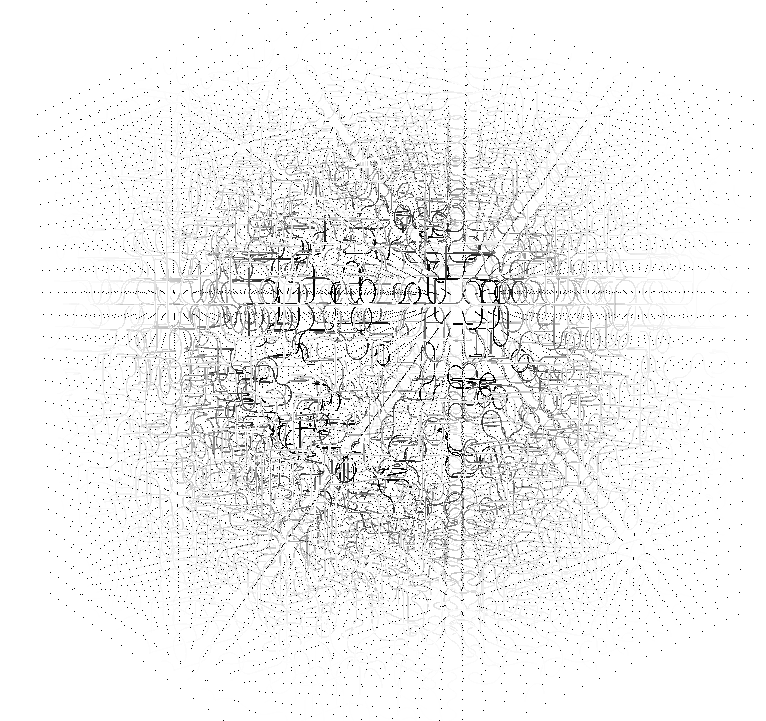}
\caption{The vortex configuration for the spherically symmetric 
inhomogeneity (\ref{J_1}) with $R=3$ and $\beta J_0=1$. 
Vortices are shown as solid lines where the coupling $J$ is maximum and 
vanish proportionally to the coupling. 
Where $\beta J\left( \vec{r} \right) \lesssim 0.1$ 
the lines become practically not visible. 
There is a central region of size $R$ where the vortex lines do not 
penetrate, and a
vortex shell around this core.}
\label{fig4}
\end{center}
\end{figure}

\begin{figure}
\begin{center}
\includegraphics[angle=0,width=0.7\textwidth]{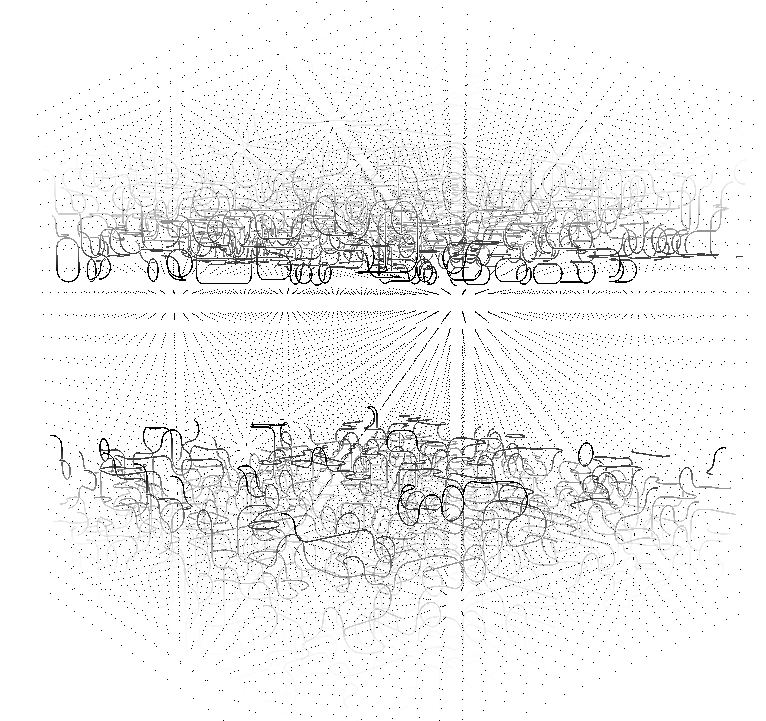}
\caption{A typical configuration with planar inhomogeneity (\ref{J_2}) 
with $\beta J_0=2.5$ and $R=2$. 
The vortex-free region size is controlled by $R$.}
\label{fig5}
\end{center}
\end{figure}

\begin{figure}
\begin{center}
\includegraphics[angle=0,width=0.7\textwidth]{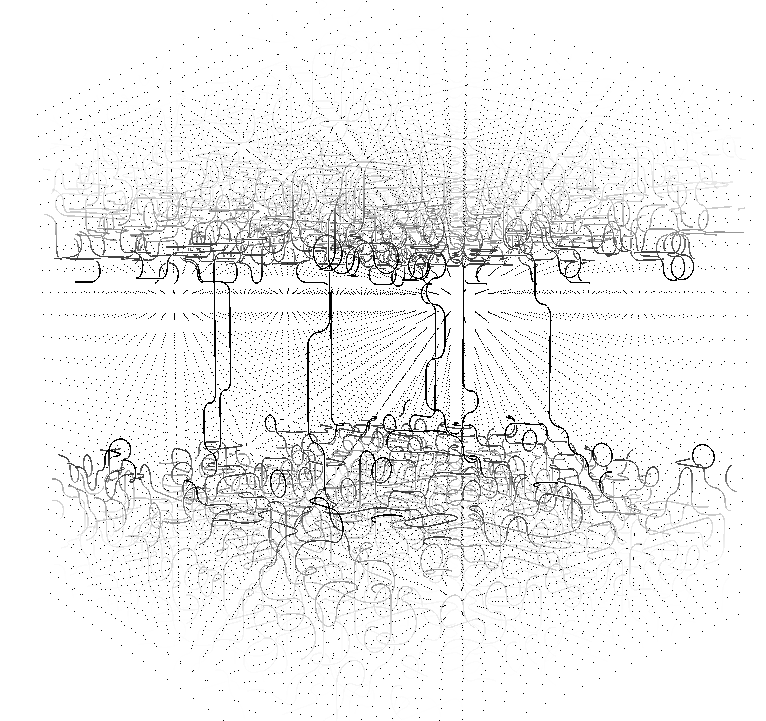}
\caption{An example of planar inhomogeneity (\ref{J_2}) 
with strong magnetic field. 
The magnetic field $\vec{{\cal B}}$ is perpendicular to the slab, with 
$\beta J_0=2.5$, $R=2$ and magnetic field $\vec{{\cal B}}=10$.}
\label{fig6}
\end{center}
\end{figure}

\begin{figure}
\begin{center}
\includegraphics[angle=0,width=0.7\textwidth]{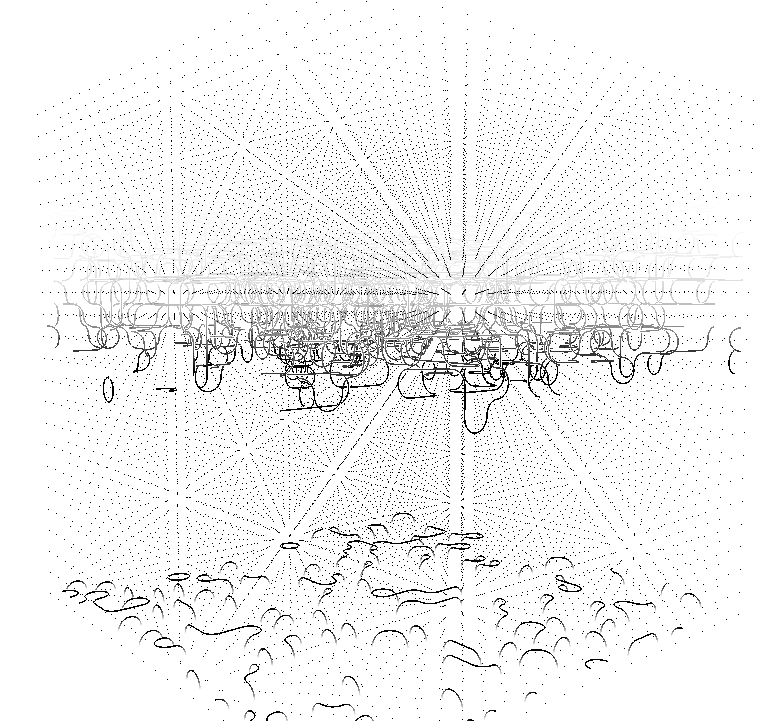}
\caption{An example configuration using the shape 
of the coupling derived from the NS crust gap 
(\ref{J_3}) with $\beta B=1.5$ and $R=4.4$.
The lower border is the inner side of the crust.}
\label{fig7}
\end{center}
\end{figure}

We examined the typical vortex configurations by taking snapshots. Figs. \ref{fig1}-\ref{fig7} 
show probable configurations in various conditions using $21^3$ sites. In Figs. \ref{fig1}-\ref{fig7} the sites 
of the cubic lattice are depicted as points and three-dimensional images of the lattice with the vortex lines are presented: 
the vortex lines, connecting the centres of the elementary cell faces, are drawn as solid lines.  
%and the different intensity of these lines 
%is just a visual aid to better distinguish the lines themselves \cite{iazzi}. 
In presence of inhomogeneous couplings, the thickness of the lines is faded proportionally to the couplings $J_{ij}$, i.e. the full solid lines correspond 
to the maximum value of the couplings and no line is drawn in absence of coupling: different intensities of grey correspond to intermediate 
couplings \cite{iazzi}.

The simulation with homogeneous coupling is shown for comparison 
and it illustrates the nature of three-dimensional $XY$ phases transition occurring at $\beta_CJ\approx0.454$, as
shown in Fig.\ref{fig1} (below $T_c$) and Fig.\ref{fig2} (above $T_c$). 
The transition point is also signalled by the behaviour of the order parameter $m=\langle cos{\phi} \rangle$ 
(not reported). Long vortex lines such as the ones in Fig. \ref{fig3} are
unstable below $T_c$: this configuration has been prepared with a sudden quench in Monte Carlo dynamics 
from very high to very low temperature ($\beta J$ quenched from $0$ to $10$). Notice that such configurations are \textit{metastable} in the 
Monte Carlo dynamics and can significantly increase the thermalization time: open lines have a very long time scale 
since they can only annihilate touching the boundaries, even if they are not very sensitive to the presence of the 
boundaries themselves. In the same way as they can survive for many simulation steps, they may take a similar amount of time 
to nucleate, so special care has been taken by varying the temperature slowly in the simulations. E.g.,  
when crossing the critical temperature, open lines can form by joining several closed loops, so that if the 
temperature starts slightly below $T_c$ the density of closed loops is high enough to allow rapid thermalization. 
In the same way, a rapid quench from above the critical temperature to a very low temperature does not allow  
the breaking of long vortex lines that can remain trapped in the system for a long time (as shown in Fig. \ref{fig3}).

In presence of an inhomogeneous coupling, regions with a higher coupling 
tend to expel vortices. In Fig. \ref{fig4} we show the effect of the 
spherically symmetric inhomogeneity (\ref{J_1}): 
the central core remains vortex free because the coupling is higher. 
Outside the core the vortex density grows rapidly and 
reaches a constant. However, as the coupling grows, 
the energy carried by vortices decreases. Eventually the 
vortex loses its meaning when the coupling goes to zero, as the phases 
$\phi$ are decoupled and only a
finite shell gives a contribution to the thermodynamic properties. 
To better illustrate this point, we decided to depict 
the vortex lines as solid 
where the coupling is maximum and make them fade proportionally to the coupling.

We also examined the case of a planar geometry defined by the Eq.(\ref{J_2}). 
The region forbidden to vortices is a slab region of size $R$ forming 
a pancake geometry: we
observe the creation of two vortex walls, as illustrated 
in Fig. \ref{fig5}.
The inner crust acts as a potential barrier for vortices: if 
the temperature in the centre is lowered, small vortex loops can
tunnel through.

The presence of a strong enough magnetic field 
can make vortices inside the barrier
as shown in Fig. \ref{fig6}. The magnetic field can be added 
in the Hamiltonian (\ref{inh_ham}) by substituting 
$\phi_i-\phi_j$ with $\phi_i-\phi_j+A_{ij}$, where $A_{ij}=\int_i^j \vec{A} 
\cdot d\vec{l}$ and the magnetic field $\vec{{\cal B}}$ is given 
by $\vec{{\cal B}}=rot \vec{A}$. 
Naturally, the lines tend to have the same winding direction. 
In the case of neutral 
particles it must be noticed that the description in terms of a
magnetic field is given in the rotating frame, 
so that the vortices are actually
rotating in the inertial reference frame.

We also investigated the case of the coupling (\ref{J_3}), 
that is derived from the NS superfluid gap. 
By fitting the results of \cite{fabrocini08} with the 
function $\left(A+Bz\right)\,e^{-Cz^2}$, we see that the gap is spatially 
asymmetric with respect to the centre of the crust. The 
parameter $B$ controls the ratio of the slopes of the gap 
at the inner and outer crust boundaries. From the fit we get that 
this ratio is $\approx 5$. This means that the inner vortex wall (bottom
side in the figure) is much thinner than the outer wall and 
it has a sharp interface with the vortex-free region as shown in 
Fig. \ref{fig7}.

Regarding the application of the Hamiltonian (\ref{inh_ham}) to rotating NS, we observe 
that using the inhomogeneity (\ref{J_3}) in presence of a strong magnetic field gives results 
qualitatively similar to those of Fig. \ref{fig6}.

\begin{figure}
\begin{center}
\includegraphics[angle=0,width=0.8\textwidth]{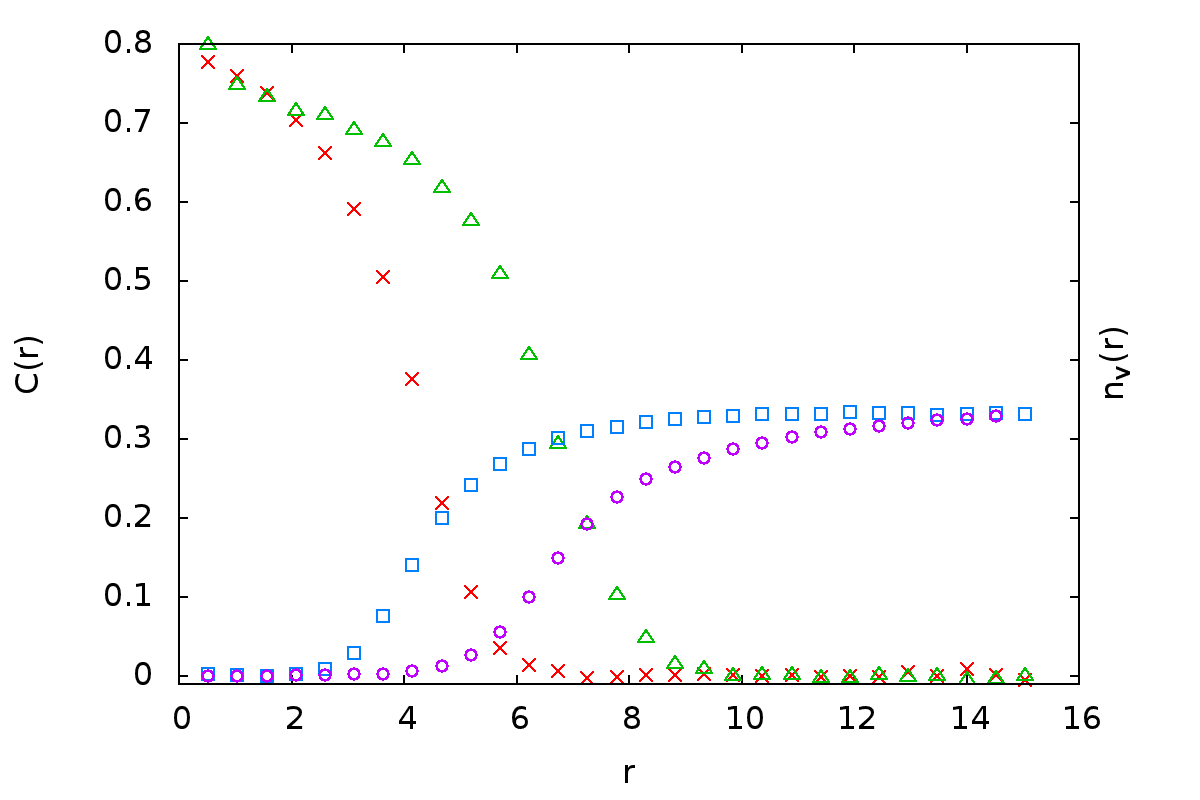}
\caption{Numerical results in the spherical geometry (Eq. (\ref{J_1}) with $\beta J_0=1$, $R=4,6$). The size of the lattice is $L=21$ and the distances are measured in units of the lattice spacing.  The correlation between the site at the origin $r=0$ and the one at distance $r$, $C(r)=\langle\cos(\phi(0)-\phi(r))\rangle$ is plotted for $R=4$ ($\times$) and $R=6$ ($\vartriangle$). The average number of vortices $n$ is also plotted ($\square$ for $R=4$, $\bigcirc$ for $R=6$) to show how vortices screen the correlation.}
\label{fig8}
\end{center}
\end{figure}

\begin{figure}
\begin{center}
\includegraphics[angle=0,width=0.8\textwidth]{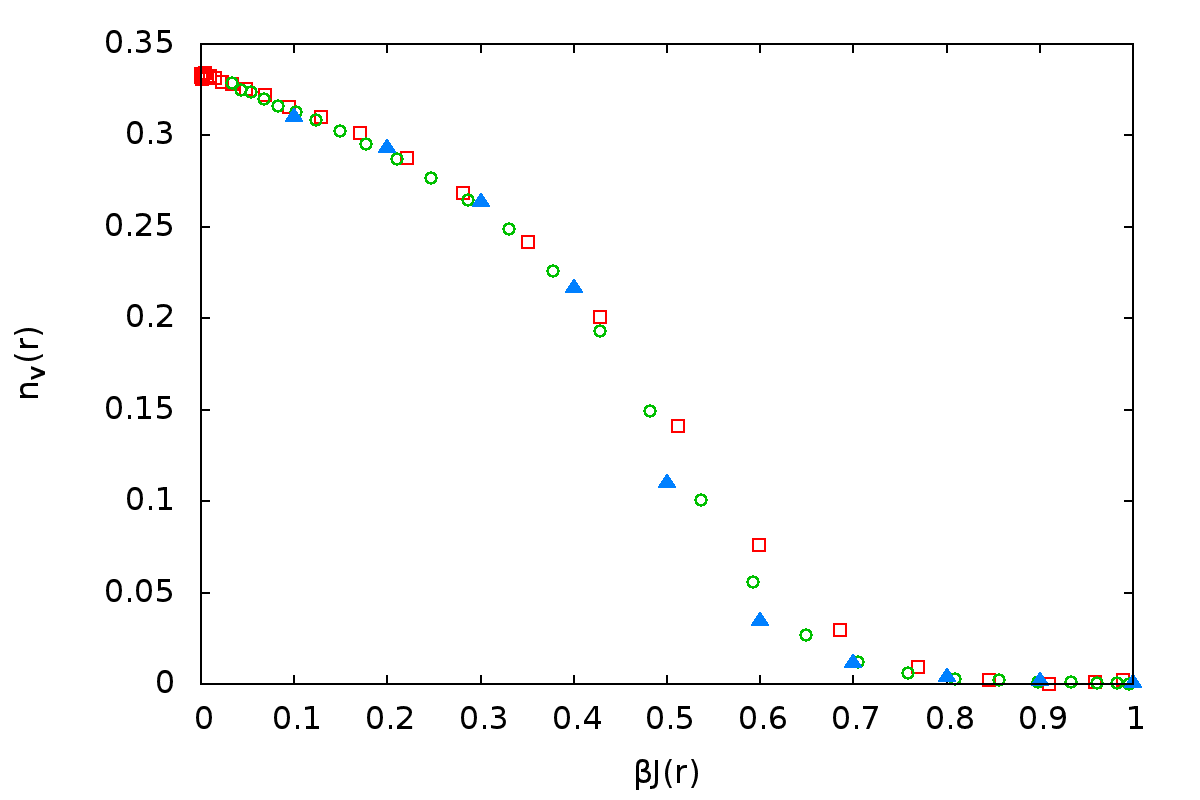}
\caption{The number of vortices as a function of the local temperature for the same model as Fig. \ref{fig8} ($\square$ for $R=4$ and $\bigcirc$ for $R=6$) and for several homogeneous systems ($\blacktriangle$ for LDA).}
\label{fig9}
\end{center}
\end{figure}

The correlation function $\langle 
\cos{\left( \phi_0 - \phi_r \right)} \rangle$ between the site at the origin 
and the site at distance $r$ is plotted in Fig. \ref{fig8}: one clearly 
sees that the correlations goes to zero in a length scale $\xi$ determined 
from the condition $\beta J(\xi) \sim 0.5$ . In the same figure we plot the 
average number of vortices at distance $r$ from the centre, which goes from 0 to a finite value at
the same length scale.

To compare our results with LDA, we calculated the average number of vortices for a homogeneous system at various temperatures and compared it to the local density of vortices as a function of the coupling strength in Fig. \ref{fig9}. The LDA agrees well with the inhomogeneous results, so it is expected that local observables are accurately described by LDA.

\begin{figure}
\begin{center}
\includegraphics[angle=0,width=0.8\textwidth]{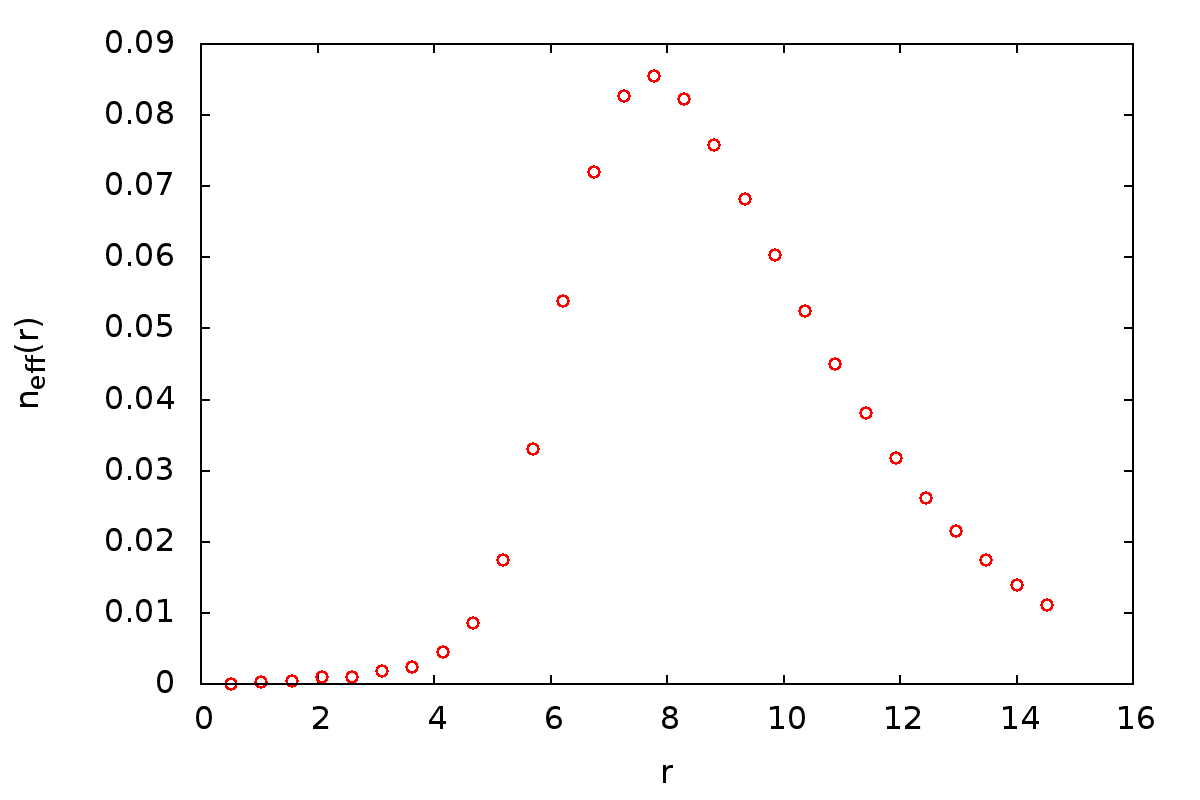}
\caption{The effective number of vortices at radius $r$ 
for the same spherical geometry and parameters as
in Fig. \ref{fig8} ($R=6$).}
\label{fig10}
\end{center}
\end{figure}

Fig. \ref{fig8} clearly illustrates that the number of vortices 
$n_v(r)$ is large in the region of small couplings 
(where the correlation with the central 
region is vanishing): therefore we introduce an effective number, $n_{eff}$ 
given by
\begin{equation}
n_{eff}(\vec{r})=\beta J(\vec{r}) \, n_v (\vec{r}).
\label{effect}  
\end{equation}
The Fig. \ref{fig10} plots the effective number of vortices (\ref{effect}) 
for the same parameters of Fig. \ref{fig8}, illustrating the anticipated 
fact that the effective number of vortices is maximum in the region with 
intermediate couplings.

\section{Conclusions}
\label{concl}

In this paper we studied the distribution of vortex lines in 
inhomogeneous three-dimensional lattices modelled by 
an $XY$ model with spatially varying coupling constants. 
The motivation of our work was two-fold and related 
to the superfluid properties of 
rotating trapped ultracold atoms in optical lattices and 
rotating neutron stars. 
Indeed the low-energy properties of ultracold atoms in deep optical lattices 
are in general described by Hubbard-like models: in the 
superfluid phase, either due to temperature 
and/or to the effect of a rotation, 
vortex lines will develop in the system. Since an external potential is 
usually present in order to trap these systems, 
then an arising issue is the determination of how the trapping potential 
affects the distribution vortex lines. 

The study the effect of inhomogeneities on the vortex lines in 
superfluids in a lattice is also relevant in the context of neutron stars. 
Theoretical calculations and indirect observational evidence 
support the idea that 
crustal free neutrons form a superfluid. Since the inhomogeneous neutron 
superfluid in the inner crust has a gap exhibiting a strong spatial 
dependence and it is experiencing 
the periodic potential created by the lattice of very neutron rich 
nuclei, then an interesting problem is the study of the effect of 
the inhomogeneity on the vortex lines in the inner crust. Also relevant 
would possibly be the individuation of a simplified model 
capturing the main properties of vortex lines in neutron stars.

We have presented results for the spatial distribution of the number of vortex lines, 
showing how the vortex lines are expelled from the region with 
higher couplings. 
The reason for this behaviour is that in the homogeneous limit 
the critical temperature $T_c$ is proportional to 
$J$ and for temperatures smaller than $T_c$ a few small 
vortex loops are thermally generated. In presence of inhomogeneity, 
there is a space dependent value of the temperature value at which the 
vortex loops start to blow out; thus, 
with the couplings varying across the lattice, 
in the regions with small coupling constants vortex loops accumulate, while
in contrast
vortices are expelled from the regions with large couplings. 
Defining an effective number of vortices to take into account 
that where the coupling is small the number of superfluid particles is also 
small, one has that the effective number of vortices will 
reach a maximum value at intermediate values of the couplings. 
Results for the correlation functions and in presence of 
a strong external magnetic field 
(or strong rotation for neutral superfluids) 
are also presented,

As a future work, it would be interesting to determine 
the vortex-vortex interaction in presence of inhomogeneities and/or 
dislocations: from this point it would be useful to write the inhomogeneous 
Hamiltonian in terms of dual variables and write the vortex-vortex 
interactions starting from the (inhomogeneous) lattice Green's functions. 
The corresponding results would then allow for the determination 
of the effective size of the region with an appreciable effective 
number of vortices. Equally interesting would also be the systematic study 
of the effects of an (eventually frustrated) external magnetic field.
  
In perspective, we think that a promising line of work is the extension of the results obtained in this paper using a simplified 
$XY$ toy model to quantitatively study vortex behaviour in rotating neutron star crusts. 
Attempts in this direction have been made using a phenomenological 
Gross-Pitaevskii equation \cite{melatos11}. 
The planned extension consists in writing an effective Hamiltonian for the superfluid neutrons put in rotation and feeling 
an effective inhomogeneous periodic potential: the superfluid dynamics will be then described by a simplified ($XY$-like) model having parameters 
determined from the underlying microscopic fermionic Hamiltonian. The resulting model could be used to make predictions for  
observable quantities and comparisons with available data, with the aim to improve the current state of the
art both in terms of modelling and of input parameters. Using numerical
results from microphysical simulations as inputs, for the pairing gap, the effective lattice potential 
and the inter-particle interaction, one could be then able to study the
behaviour of vortices in a realistic way and hopefully bridge
the gap between the astrophysical vast dataset and the many-body theoretical
picture describing the crust. 

{\bf Acknowledgements} We thank J. Miller, P. Sodano, M. Rasetti and 
S. R. Shenoy for many interesting discussions. K.S. was supported by NSF grants PHY0757703 and PHY1067777. He thanks Luciano Reatto for
guidance, and many stimulating discussions throughout his career. S.F. would like to thank Luciano Reatto, 
first, for having introduced him to the Quantum Monte Carlo world, then for the always 
exciting and illuminating collaborations, and, finally, for the never ending friendship.

%\begin{figure}\label{}
% \includegraphics[angle=270,width=1.1\textwidth]{.eps}
%\caption{}
%\end{figure}

% \appendix
% \section{}

\end{document}